\documentclass[sn-mathphys,Numbered]{sn-jnl}
\usepackage{natbib}
\usepackage[utf8]{inputenc}
\usepackage[english]{babel}
\usepackage{graphicx}%
\usepackage{multirow}%
\usepackage{amsmath,amssymb,amsfonts}%
\usepackage{amsthm}%
\usepackage{mathrsfs}%
\usepackage[title]{appendix}%
\usepackage{xcolor}%
\usepackage{textcomp}%
\usepackage{manyfoot}%
\usepackage{booktabs}%
\usepackage{algorithm}%
\usepackage{algorithmicx}%
\usepackage{algpseudocode}%
\usepackage[normalem]{ulem}
\usepackage{listings}%
\setcitestyle{open={[},close={]}}
\usepackage{ragged2e}
\usepackage{todonotes}

\usepackage{multicol}

\begin{document}


\title[ Equivalent Gravities and   Equivalence Principle: Foundations and experimental implications]{ Equivalent Gravities and Equivalence Principle: Foundations and experimental implications}

\author[1,2]{\fnm{Christian} \sur{Mancini}}\email{christian.mancini@na.infn.it}
\author[3,4,5]{\fnm{Guglielmo Maria} \sur{Tino}}\email{guglielmo.tino@unifi.it}
\author[1,2,6]{\fnm{Salvatore} \sur{Capozziello}}\email{capozziello na.infn.it}

\affil[1]{\orgdiv{Scuola Superiore Meridionale}, \orgaddress{\street{Largo San Marcellino 10}, \postcode{I-80138}, \state{Napoli}, \country{Italy}}}

\affil[2]{\orgdiv{Istituto Nazionale di Fisica Nucleare, Sezione di Napoli}, \orgaddress{\street{Complesso Universitario di Monte S. Angelo, Via Cinthia Edificio 6}, \postcode{I-80126}, \state{Napoli}, \country{Italy}}}

\affil[3]{\orgdiv{Department of Physics and Astronomy}, \orgname{Università di Firenze}, \orgaddress{\street{via Sansone 1}, \postcode{I-50019}, \state{Sesto
Fiorentino}, \country{Italy}}}
\affil[4]{\orgdiv{European Laboratory for Non-Linear Spectroscopy (LENS)}, \orgaddress{\street{via Nello Carrara 1}, \postcode{I-50019}, \state{Sesto
Fiorentino}, \country{Italy}}}
\affil[5]{\orgdiv{Istituto Nazionale di Fisica Nucleare, Sezione di Firenze}, \orgaddress{\street{via Sansone 1}, \postcode{I-50019}, \state{Sesto
Fiorentino}, \country{Italy}}}

\affil[6]{\orgdiv{Dipartimento di Fisica "E. Pancini"}, \orgname{Università di Napoli “Federico II”}, \orgaddress{\street{Via Cinthia Edificio 6}, \postcode{I-80126}, \state{Napoli}, \country{Italy}}}

\abstract{The so-called Geometric Trinity of Gravity includes General Relativity (GR), based on spacetime curvature; the Teleparallel Equivalent of GR (TEGR), which relies on spacetime torsion; and the Symmetric Teleparallel Equivalent of GR (STEGR), grounded in nonmetricity. Recent studies demonstrate that GR, TEGR, and STEGR are dynamically equivalent, raising questions about the fundamental structure of spacetime, the under-determination of these theories, and whether empirical distinctions among them are possible.
The aim of this work is to show that they are equivalent in many features but not exactly in everything. In particular, their relationship with the Equivalence Principle (EP) is different. The EP is a deeply theory-laden assumption, which is assumed as fundamental in constructing GR, with significant implications for our understanding of spacetime. However, it introduces unresolved conceptual issues, including its impact on the nature of the metric and connection, its meaning at the quantum level, tensions with other fundamental interactions and new physics, and its role in dark matter and dark energy problems.
In contrast, TEGR and STEGR recover the EP, in particular in its strong formulation, but do not rely on it as a foundational principle. The fact that GR, TEGR, and STEGR are equivalent in non-trivial predictions, but the EP is not necessary for TEGR and STEGR, suggests that it may not be a fundamental feature but an emergent one, potentially marking differences in the empirical content of the three theories.
Thus, the developments within the Geometric Trinity framework challenge traditional assumptions about spacetime and may help to better understand some of the unresolved foundational difficulties related to the EP.}

\keywords{Theories of Gravity, Equivalence Principle, Spacetime Structure\\ \\\textbf{Foundations of Physics} (2025), 55:69\\
https://doi.org/10.1007/s10701-025-00882-x \\Received: 19 January 2025 / Accepted: 28 July 2025}



\maketitle
\tableofcontents

\section{Introduction}\label{Introduction}
Despite General Relativity (GR) being considered the \lq\lq standard theory" of gravity, there has never been a period in its history without serious alternatives being proposed and developed, with different approaches and motivations \citep{capozziello2011extended, NOJIRI20171, smeenk2023dark, Will2018theory}. Particularly in the last four decades, in the so-called \lq\lq precision cosmology era", advancements in cosmological observations, gravitational wave physics and high precision tests have highlighted significant shortcomings, both at theoretical and observational level, both at small (UV) scales, and at large (IR)  scales. \citep{capozziello2011extended}.

Several authors find it natural and elegant to consider improving the gravitational components of  field equations \citep{capozziello2011extended, NOJIRI20171, smeenk2023dark}, by exploring extended or alternative theories of gravity. There are many types of approaches to modified gravity theories \citep{capozziello2011extended, NOJIRI20171}: straightforward extensions like $f(\mathcal{R})$ gravity, Vector-Scalar-Tensor theories, where geometry can non-minimally couple to new fields; higher-order theories, where derivatives of  metric components higher than second order can appear; theories with modified geometry; theories based on different principles, such as MOdified Newtonian Dynamics (MOND); unifying theories attempting to quantize gravity, and so on. 

A general approach  can be  inserted in the context of the so-called metric-affine theories of gravity, which fall into the category of approaches  where geometry is enlarged and improved \citep{capozziello2022comparing}. In particular, in 1919 Palatini showed that the metric tensor and the affine connection, which constitute GR, can be considered as two different geometric structures and can be varied independently \citep{ferraris1982variational}. These considerations led to the development of theories where the field equations can be formulated in terms of other geometric invariants: the torsion tensor and the non-metricity tensor. Together with the curvature tensor, considering these three geometric objects, we can build up different  theories, where GR is a particular case in a lake of more general metric-affine theories, where torsion and non-metricity are set to zero. Among these theories, of particular interest is the so-called Geometric Trinity of Gravity \citep{beltran2019geometrical, bahamonde2023teleparallel, capozziello2022comparing, Capozziello2023boundary}, which comprises GR, built upon the metric tensor and grounded on the curvature of spacetime; the Teleparallel Equivalent of GR (TEGR), formulated in terms of torsion of spacetime and relying on tetrads and spin connection; and the Symmetric Teleparallel Equivalent of GR (STEGR), built on nonmetricity and constructed from metric tensor and affine connection. For this reason, in this case we can speak of \textit{modified spacetime} rather than modified gravity, as some authors suggested \citep{martens2020dark}.

Significantly, these three theories have been found to be dynamically equivalent to GR, as the names suggest, since their actions differ by boundary terms, which are dynamically irrelevant because the spacetime is asymptotically flat at infinity, and thus these terms are vanishing \citep{capozziello2022comparing}. The Geometric Trinity recently gained a lot of attention in both the theoretical \citep{beltran2019geometrical,bahamonde2023teleparallel,capozziello2022comparing,
Capozziello2023boundary,capozziello2024equivalence} and the philosophical literature \citep{mulder2023spacetime,march2023geometric,wolf2023underdetermination,chen2024equivalence, Durr2024}. From these recent theoretical developments, many questions arise. Should we consider TEGR and STEGR as proper alternative theories to GR or merely different dynamical formulations? Is it possible to empirically discriminate among them? If they are dynamically equivalent, is gravitation  given by curvature, torsion, or non-metricity of spacetime at some fundamental level? Some authors have argued that TEGR is empirically equivalent to GR, having the same dynamics - that is, the same action up to a boundary term \citep{knox2011newton, weatherall2025some}. However, as shown by other authors, the empirical content of a theory is not exhausted by its dynamics, i.e. by the equations of motion \citep{Wolf_2023boundaries}. Therefore, two theories with the same action may still exhibit differences in their empirical content, as we shall argue.

In this work, it will be shown that a closer inspection on the equivalent features reveals  crucial differences among the theories, that is, their relation with respect to the Equivalence Principle (EP). The EP was one of the most important assumptions that led to the discovery of GR, famously described as the \lq\lq midwife" that helped Einstein to develop the theory \citep{Synge1960-SYNRTG, lehmkuhl2021equivalence}. As we will see in detail, there are many open conceptual difficulties, which are direct or indirect consequences of the EP imposition at the foundation of the theory. Some of them puzzled Einstein himself until the very end of his life, such as the coincidence between the geodesic and the causal structure, or the fact that the fundamental object of the theory is the Riemannian metric $g_{\mu\nu}$ instead of the connection $\Gamma^{\rho}_{\mu \nu}$. The fact that the EP is such a theory-laden principle has led some authors to describe it as a \lq\lq beast" \citep{lehmkuhl2021equivalence}, or a bunch of beasts, given all its different formulations.

Significantly, the Geometric Trinity suggests that there are viable theories of gravity which need not impose the EP. The fact that these three representations of gravity are dynamically equivalent, and that the EP can be recovered but not at the foundation of TEGR and STEGR, suggests that the EP could be not a fundamental principle, but an emergent feature related to some symmetry or gauge \cite{capozziello2024equivalence,Capozziello:2024clc,Bajardi:2022ypn}. Therefore, it will be argued that, at some level, the EP might constitute a direct difference in the empirical content between the three representations and could allow us to discriminate among them.

This result is relevant because, if it is the case, it could allow one to relax this theory-laden assumption at fundamental level and address some of the open problems. This, in some sense, is in line with the intuition by Synge, who considered it only a midwife and not a fundamental feature of the world \citep{Synge1960-SYNRTG}.

In the EP discussion, we follow the approach by Lehmkuhl \citep{lehmkuhl2021equivalence}, as well as the recent contribution by Read and Teh on the EP in teleparallel gravity \citep{Read2023}. We adopt their framework and extend the discussion to the full Geometric Trinity.\\

The paper is organized as follows. In Sec.~$\ref{General Relativity assumptions and difficulties}$, the open fundamental difficulties entailed by the EP are discussed. In Sec. \ref{Equivalent Gravities},  recent results on the equivalent features in the Geometric Trinity of Gravity, relevant for this work,  are presented. Then, in Sec.\ref{Epistemological considerations on Equivalent Gravities}, the relation between the EP and the Geometric Trinity will be analysed. In Sec.\ref{Discussion}, the epistemological and experimental implications will be discussed.
In Sec.~$\ref{Conclusions}$, conclusions are drawn.

\section{General Relativity: assumptions and  shortcomings}
\label{General Relativity assumptions and difficulties}
There are many significant conceptual difficulties that emerge from the assumption and the application of the EP, which are known in the physical and foundational literature, and which are sufficient to give to the EP the reputation of being \lq\lq a beast" \citep{lehmkuhl2021equivalence}.

Without entering in historical details, GR is built upon the EP, as well as upon other fundamental assumptions. In the following, we adopt the Lehmkuhl framework \citep{lehmkuhl2021equivalence}, and we follow, in particular, the definitions provided by Read and Teh in a recent discussion of the EP in TEGR \citep{Read2023}.

In its weaker form (WEP), the EP states that the gravitational mass of a body is equal to its inertial mass (called WEP2 in Ref. \citep[p. 5]{lehmkuhl2021equivalence} and \citep[p. 3484]{Read2023}). The empirical indistinguishability between gravitational and inertial effects was a crucial starting point for Einstein in formulating his version of the EP, which is now known as the Einstein Equivalence Principle (EEP) \citep[p. 10]{lehmkuhl2021equivalence}:
\begin{quote}
    \normalsize{\textbf{EEP:} Gravity and inertia are the same in their very essence ("wesensgleich").}
\end{quote}
\begin{justify}
Mathematically, it means that both gravitational and inertial effects are represented by the components of the same compatible connection \citep[p. 3486]{Read2023}.

Finally, the link between the EEP and the local validity of Special Relativity (SR), first underlined by Pauli in 1921 \citep[pp. 21-22]{lehmkuhl2021equivalence}, constitutes the basis for the strong formulation of  EP (SEP). According to the definition by Pauli, for every infinitely small region of the world, there always exists a coordinate system such that gravitational effects can be neglected. Einstein did not call it as the SEP, but he considered the EEP and the local validity of SR as two strictly related concepts \citep{lehmkuhl2021equivalence}. Read and Teh state the SEP as follows \citep[p. 3486]{Read2023}:
\end{justify}
\begin{quote}
    \normalsize{$\mathbf{SEP_{1,EEP}:}$ At any point $p~ \in~ M$, a manifold, one can find an orthonormal  frame in which gravito-inertial effects, as represented by connection coefficients, vanish.}
\end{quote}
Then, Read and Teh define a version of the SEP which do not assume the EEP, which will be useful in relation to TEGR:
\begin{quote}
\normalsize{$\mathbf{SEP_{1,\neg EEP}:}$ At any point $p~ \in~ M$, a manifold, one can find a frame in which inertial effects, as represented by connection coefficients, cancel gravitational effects, as represented by some tensor quantity.}
\end{quote}
Adding that the frames in which the principle holds are related by Lorentz transformations, we arrive to what Read and Teh call $SEP_2$ \citep[p. 3487]{Read2023}:
\begin{quote}
\normalsize{$\mathbf{SEP_{2, EEP}:}$  SEP\(_{1,\mathrm{EEP}} \) holds, and the frames in which this principle holds are related by Lorentz transformations.\\
$\mathbf{SEP_{2,\neg EEP}:}$ SEP\(_{1,\mathrm{\neg EEP}} \) holds, and the frames in which this principle holds are related by Lorentz transformations.}
\end{quote}

The standard frame where the physics of SR is locally recovered, the metric tensor $g_{\mu\nu}$ takes the Minkowski form $\eta_{\mu\nu}$, and the connection coefficients $\Gamma^{\rho}_{\mu \nu}$ vanish, is called the Local Inertial Frame (LIF). The requirement that such a frame exists at every point $p~ \in~ M$, demands the metric compatibility condition $\nabla_{\lambda}g_{\mu\nu}=0$ \citep[p. 313]{misner1973gravitation}.

Note that there are also other different formulations of the SEP, as the extension of the EEP also to bodies with non-negligible self-gravitational interactions and gravitational experiments \citep[p. 76]{Will2018theory}. However, the relation between these different definitions is not yet clear and this issue is not addressed in this work.

In the following, we will list eight foundational problems relevant for this work.

\paragraph{i) Coincidence of the causal and the geodesic structure}
As a consequence of the imposition of the SEP, the Christoffel symbols $\Gamma^{\rho}_{\mu \nu}$ coincide with the Levi-Civita connection. Assuming the SEP, the unique possible affine symmetric and metric compatible connection is the Levi-Civita one. The \textit{a priori} assumption of the SEP in this way selects the Levi-Civita connection \citep[p. 314]{misner1973gravitation}:
\begin{equation} \label{eq:affine_connections2}
\begin{aligned}
& g_{\mu\nu}(p)=\eta_{\mu\nu}, \quad \Gamma^{\rho}_{\mu \nu}(p)=0, \quad \nabla_{\lambda}g_{\mu\nu}=0, \quad \Gamma^{\rho}_{\mu \nu}=\Gamma^{\rho}_{\nu \mu}
\end{aligned}
\end{equation}

\centering
$ \Downarrow $
\begin{equation} \label{eq:affine_connections3}
\begin{aligned}
& \Gamma^{\rho}_{\mu \gamma} =
\begin{Bmatrix}
\rho \,\\
\mu\nu
\end{Bmatrix}
= \frac{1}{2} g^{\rho \gamma} \left( \partial_{\mu} g_{\lambda \nu} + \partial_{\nu} g_{\mu \lambda} - \partial_{\lambda} g_{\mu \nu} \right)
\end{aligned}
\end{equation}

\justifying
Consequently, by construction, the Levi-Civita connection has no indipendent dynamics, but it is a by-product of the metric $g_{\mu\nu}$, containing its derivatives. Physically, it represents the apparent forces acting on the body due to the curved geometric background. This means that the metric $g_{\mu\nu}$ determines, at the same time, the causal structure (light cones with rods and clocks) and the geodesic structure (the free fall of test particles) \citep{misner1973gravitation, tino2020precision}. However, it is important to underline that {\it a priori} there is no relation between the connection $\Gamma^{\rho}_{\mu \nu}$ and the metric tensor $g_{\mu \nu}$, but it is a consequence of the imposition of  SEP. In fact, this coincidence does not work anymore for extensions of GR as $f(\mathcal{R})$ \citep{Allemandi2006}.  This unjustified coincidence is a first conceptual problem, which has been widely discussed in the literature (see e.g. \citep{capozziello2009dark}).

\paragraph{ii) The metric as the fundamental object of the theory}
The second problematic direct consequence of this picture is the following. As mentioned, in GR the truly fundamental dynamical object is the metric tensor $g_{\mu\nu}$. However, there are reasons not to regard it as the gravitational field, since it represents a set of potentials. According to Einstein himself, the proper gravitational field should be identified with the connection components \citep{einstein1916friedrich, LEHMKUHL200883, Janssen2007, Fletcher_2025}, as they are derivatives of the metric and represent a straightforward generalization of the Newtonian gravitational field, which is the gradient of the potential \citep[p. 846]{Janssen2007}. The connection components also appear in the geodesic equation, which Einstein considered the successor and generalization of the Newtonian law of inertia \citep[p. 15]{lehmkuhl2021equivalence}, thus making them natural candidates for representing the gravitational field:
\begin{align} \label{eq:geodesics}
\frac{d^2x^{\rho}}{d\tau ^2} +\Gamma^{\rho}_{\mu \gamma} \frac{dx^{\mu}}{d\tau} \frac{dx^{\mu}}{d\tau}=0\,,
\end{align}
where $\Gamma^{\rho}_{\mu \gamma} \frac{dx^{\mu}}{d\tau} \frac{dx^{\mu}}{d\tau}$ represents the generalization of the Newtonian forces. For a discussion of other candidates for the gravitational field, see Ref. \citep{LEHMKUHL200883}.

This problem puzzled Einstein until the very end of his life, when he remarked again that the fundamental element should be the connection, and only indirectly the Riemannian metric $g_{\mu\nu}$ \citep[pp. XVIII-XIX]{pantaleo1955cinquant} (see also \citep[p. 9]{goenner2014history}). Moreover, in experiments, what we really measure are the forces (or the accelerations), which are represented by the connection $\Gamma^{\rho}_{\mu \nu}$. And as we have seen, the connections are the first $derivatives$ of the metric and the second derivatives of the local inertial coordinates:

\begin{align} \label{eq:second_derivatives}
\Gamma^{\rho}_{\mu \gamma} = \frac{dx^{\rho}}{d\xi^{\sigma}} \frac{d^2\xi^{\sigma}}{dx^{\mu}dx^{\gamma}}\,.
\end{align}

Therefore, the quantity we observe in typical experiments is not related to the metric but to the connection, which in GR gains its dynamics from the former. 

There were also other reasons for Einstein to not consider the metric $g_{\mu\nu}$ as the fundamental object, as the fact that it seemed to him too similar to the concept of Newtonian absolute space, the overcoming of which was one of his first objectives \citep{lehmkuhl2021equivalence}.

Therefore, thanks to the SEP, spacetime is described by the double $\langle M,g_{\mu\nu}\rangle$, i.e. the Riemannian manifold, where $M$ is the manifold and $g_{\mu\nu}$ the metric tensor, that is the fundamental object.

\paragraph{iii) Tensions with new physics predictions}
As a general consideration, it is possible to state that   \lq\lq new physics" naturally predicts the violation of the  SEP at some level. The simplest case is in Scalar-Tensor Theories, where, in the Einstein Frame, the mediation of the \lq\lq fifth force" causes a difference in the free fall between different objects \citep{capozziello2011extended}, which violates the WEP, and therefore the SEP. Then, in theories featuring Quintessence, where the cosmological constant is replaced by a slowly evolving scalar field, one expects that the coupling of this field with matter induces gravitational forces that depend on the composition of the body. This clearly violates the WEP. Scalar fields violating the WEP are also predicted by theories involving extra dimensions, such as String Theory (ST). As argued in Ref.\citep{damour2012theoretical}, current precision levels of WEP tests (today $10^{-15}$, as we will see in Sec. \ref{Discussion}) should not discourage further research, as ST could imply WEP violations even further. For this reason, in Ref.  \citep{damour2012theoretical}, it is  suggests that the WEP experiments are the most sensitive tools that we have to test  new physics. Similarly, some authors argue that  quantum properties of gravity could have observable experimental consequences at low energies, such as the dependence of geodesic motion on the mass of  test particles, as explored in \citep{fortier2007precision, lammerzahl2009determines}. Other authors \citep{blasone2019equivalence} derive a direct violation of  WEP at finite temperature from Quantum Field Theory. Then, famously, there is MOND (MOdified Newtonian Dynamics), which from observational constraints, should induce violations of  the SEP \citep{milgrom2019noncovariance}. Data from open clusters show a smaller mass discrepancy than would be required if MOND obey the SEP, even though it seems that these predictions consider the definition of  SEP as the extension of  EEP also to bodies with non-negligible self-gravitational interactions \citep[p. 76]{Will2018theory}, which is not the same concept under consideration in this work. Finally, it can be shown that MOND is just a particular case of some field theory (e.g. $f(\mathcal{R})$ gravity \cite{Vesna,Tula}) so SEP can be questioned as soon as one relaxes the hypothesis that GR is the \lq\lq only" viable theory of gravity.

On this line, in Ref.\citep{damour2012theoretical}, it is suggested that any bias towards metric theories is entirely unjustified, both historically and from the perspective of contemporary fundamental physics.

\paragraph{iv) Validity of the SEP at quantum level}
A further significant issue is that we do not know if the WEP, and therefore the SEP is valid at quantum level. At the moment,  we are only assuming its validity. We are not even sure if this principle could be generalized by the quantum formalism. There are attempts in this direction, but there are conflicting opinions among physicists \citep{giacomini2020einstein, tino2020precision, Torrieri:2022znj}. Since at quantum level particles behave like wave packets, it is difficult to make sense of the concepts  of free fall universality or the identity between the gravitational and the inertial mass. Anyway, given quantum mechanics, we have no a priori reasons to postulate the WEP validity.

\paragraph{v) Dark Energy and Dark Matter as possible geometric issues}
There is also the problem of Dark Energy (DE) and Cold Dark Matter (CDM), on which there is a big debate in the community. As already mentioned, several authors find it more elegant to consider altering the gravitational component of the field equations \citep{capozziello2011extended} in order to address the dark phenomenology as a geometric problem, instead of a fluid one \cite{Capozziello:2006uv}. The main practical reason is that, up to today, there is no final indication that dark side components could be addressed by new fundamental particle sector \cite{Bidin:2010rj,Bidin:2011zrx,Serpico:2012ey,Serpico:2021cnm}. This debate between the dark fluid hypothesis and modified gravity approaches is impressively widespread in the community of philosophers of physics \citep{martens2023doing, smeenk2023dark, duerr2023methodological, kashyap2023general}, with particular focus on the metric postulate. Authors argued that since any viable metric theory of gravity finds dark matter in the sieve \citep{kosso2013evidence}, it could be the case that a non-metric theory of gravity could help in better understanding this problem. 

The interesting thing is that, similarly to GR, where we can extend it to $f(\mathcal{R})$ gravity, $f(T)$ and $f(Q)$ gravity are the extensions of TEGR and STEGR, respectively. Where $R$ is the Ricci curvature scalar, $T$ is the torsion scalar and $Q$ the non-metricity scalar, while $f(\mathcal{R})$, $f(T)$ and $f(Q)$ are more general functions of them. The dynamical equivalence in the Geometric Trinity holds only for theories linear in the scalar invariants and not for the extensions, for different reasons \citep{cai2016f, capozziello2022comparing, Capozziello2023boundary}. Firstly, the extensions give rise to dynamics with different degrees of freedom. In particular, in $f(\mathcal{R})$ gravity, we have field equations of fourth order, in metric representation, whereas $f(T)$ and $f(Q)$ still remains of second-order. In addition, in $f(T)$ and $f(Q)$, we cannot choose, in general, a gauge to simplify the calculations, as in the cases of TEGR and STEGR. The point is that similarly to the fact that $f(\mathcal{R})$ theories are being studied to resolve shortcomings of GR at different scales (see \citep{capozziello2011extended}), also the extensions of TEGR and STEGR  show interesting features in this direction. Physicists are already exploring $f(T)$ and $f(Q)$ gravities to study not only DE but also large structures, bouncing cosmologies, quantum cosmology, relativistic MOND theories, cosmography, inflation and gravitational waves (see for instance \citep{cai2016f, heisenberg2024review}).

Therefore, the exploration of these alternatives to GR may provide a promising path to a successor and more fundamental theory. We can say \lq\lq more fundamental" because TEGR and STEGR have not to postulate the EP in any form, as we will see,  so they could be regarded as generalizations of GR, as Einstein already guessed.

\paragraph{vi) Difference with other fundamental interactions}
The  SEP sets gravity apart from other fundamental interactions of Nature. In today's theoretical physics, it is believed to be very important to formulate theories as $gauge$ theories, since it works so well with other fundamental interactions \citep{blagojevic2001gravitation}. The interesting fact is that gravitation can be reformulated as a gauge theory properly without assuming the EP in any form \citep{capozziello2022comparing}. See Ref. \citep{weatherall2025some, March2025equivalence, wallace2015fields, duerr2025clarifying, read2025clarifying} for philosophical discussions on TEGR and STEGR as gauge theories.

\paragraph{vii) Epistemic justification of the coincidence between $m_G$ and $m_I$}
There is the foundational problem of how we can justify the  coincidence between  gravitational  and  inertial mass, which is the basis of the formulation of the WEP. For Newton, the mass of any body, understood as the property of the body itself to respond to a force, corresponded to its \lq\lq weight" , which is its property to respond to gravity. In modern terms, we would say that inertial mass $m_I$ is equal to passive gravitational mass $m_G$, terms coined by Bondi \citep{bondi1957negative}. Einstein said that it was precisely the famous E\"otv\"os experiments on the equivalence between $m_I$ and $m_G$ that directly inspired him in the formulation of the EEP \citep{lehmkuhl2021equivalence} and which, in fact, constitutes one of its cornerstone. 

Today, for many people, this equivalence might seem obvious, but back then, it was not, and it would not be even today if we \lq\lq forgot" to acknowledge this principle. In other words, {\it prima facie}, there are no reasons to postulate the identity $m_G\equiv m_I$, and it is not related to some fundamental symmetry. Einstein himself embraced it from empirical reasons. So apart from observations, how one could even imagine this equivalence? 

\paragraph{viii) Curvature over torsion of spacetime}
It is important to underline that in GR, torsion of spacetime is set to zero a priori, since with the imposition of the SEP, Einstein chose the symmetric connection, the Levi-Civita one, and so curvature of spacetime. In 1922, Cartan explored a different direction, considering a natural extension of GR constituted not only by the Levi-Civita connection, but also by the torsion tensor, that is the antisymmetric part of a metric compatible affine connection. In this way, he developed a geometric formulation where he suggested that torsion can be physically related to the intrinsic (quantum) angular momentum of matter and it vanishes in vacuum \citep{capozziello2022comparing}. Einstein himself, in the period 1923 – 1933, tried different geometries for the construction of a unified field theory \citep[p. 57]{goenner2014history}. In 1928, he published his first paper on “fernparallelism”, or teleparallelism \citep{einstein2005riemann}. As we will see in more detail, since TEGR is found to be dynamically equivalent to GR, classical tests that were understood to confirm the curvature of spacetime can similarly be understood as confirming the torsion of spacetime. This is known as the problem of \textit{geometric under-determination} \citep{mulder2023spacetime, wolf2023underdetermination}.

A priori, why prefer curvature over torsion of spacetime? It seems natural to our minds to think of a massive object as causing curvature of spacetime. However, {\it prima facie}, it is natural, in a similar way, to think that a massive object could also cause torsion. Consider, for instance, a rotating black hole. Therefore, curvature of spacetime is not, {\it a priori}, more probable than torsion. Similar considerations can be applied also to non-metricity, but curvature and torsion are intuitively easier to think about metaphysically.

So GR assumes a priori the SEP, and therefore curvature. The Geometric Trinity challenges also this fundamental assumption on spacetime focusing only on equivalence of dynamics.\\

In conclusion, there are many conceptual difficulties as direct or indirect consequences of the EP. As it will be argued in the next sections, the framework of metric-affine theories and the relaxation of the assumption of the EP could help in addressing these issues, apparently maintaining the same consolidated successes.

\section{Equivalent Gravities}\label{Equivalent Gravities}
We will now summarize the main achievements of Geometric Trinity which are relevant for the present discussion. We refer to some recent works \citep{capozziello2022comparing, capozziello2024equivalence, krvsvsak2019teleparallel, beltran2019geometrical}.\\

As we have previously seen, the GR spacetime is assigned by the double 
\begin{equation}
\langle M,g_{\mu\nu}\rangle\,,
\end{equation}
due to the imposition of the  SEP. On the contrary, following the Palatini approach \citep{palatini1919deduzione}, 
the metric $g_{\mu\nu}$ and the connection $\Gamma^{\rho}_{\mu \nu}$ can be varied independently. In this case, spacetime is assigned by  the triple:
\begin{align} \label{eq:variety}
    \langle M,g_{\mu\nu}, \Gamma^{\rho}_{\mu \nu}\rangle, 
\end{align}
where $g_{\mu\nu}$ determines the causal structure while the connection $\Gamma^{\rho}_{\mu \nu}$ determines the free fall \citep{beltran2019geometrical}. 

\begin{figure}
    \centering
    \includegraphics[height=3.5cm]{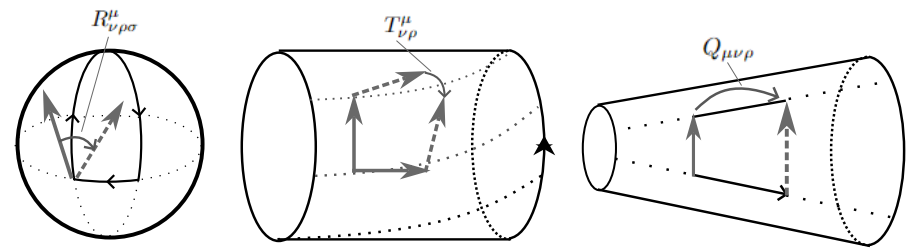}
    \caption{Typical way of defining the three geometrical invariants \citep{bahamonde2023teleparallel, capozziello2022comparing}. There is curvature of spacetime if after having paralleling transported a vector along a closed loop, there is a non-zero angle between the initial and final vectors. There is torsion of spacetime if by paralleling transporting two vectors one along the other, it is not possible to close the parallelogram. Finally, non-metricity occurs if there is a change in the length of the vector when it is moved along a curve.}
    \label{fig:Fig1}
\end{figure}

Einstein himself recognized as significant the Palatini method, since it represents a simplification of the relativistic formalism \citep[p. XXIII]{pantaleo1955cinquant}, or a generalization, we would say. With the Palatini approach, the connection $\Gamma^{\rho}_{\mu \nu}$ can be written in a more general form considering the \textit{affine} connection \citep{beltran2019geometrical, bahamonde2023teleparallel}:
\begin{align} \label{eq:general_connection}
	\Gamma^{\rho}_{\mu \nu} = \begin{Bmatrix}
\rho \\
\mu\nu
\end{Bmatrix} + K^{\rho}_{\mu \nu} + L^{\rho}_{\mu \nu}. 
\end{align}
Where $\begin{Bmatrix}
\rho \\
\mu\nu
\end{Bmatrix}$ is the Levi-Civita connection, $K^{\rho}_{\mu \nu}$ and $L^{\rho}_{\mu \nu}$ are the contortion and the distortion tensors, respectively \citep{bahamonde2023teleparallel}:
\begin{align} \label{eq:HE_action}
K^{\rho}_{\mu \nu}=\frac{1}{2} (T_{\mu\nu}^{\rho}+T_{\nu\mu}^{\rho}-T^{\rho}_{\mu \nu})\\
L^{\rho}_{\mu \nu}=\frac{1}{2} (Q^{\rho}_{\mu \nu}-Q_{\mu\nu}^{\rho}-Q_{\nu\mu}^{\rho}).
\end{align}
$T^{\mu}_{\nu \rho}$ and $Q_{\mu \nu \rho}$ are the torsion and the non-metricity tensors, and $R^{\mu}_{\nu \rho \sigma}$ is the  curvature tensor  \citep{bahamonde2023teleparallel}:
\begin{align} \label{eq:curvature}
	R^{\mu}_{\nu \rho \sigma} = \partial_{\rho}\Gamma^{\mu}_{\nu \sigma}-\partial_{\sigma}\Gamma^{\mu}_{\nu \rho}+\Gamma^{\mu}_{\tau \rho} \Gamma^{\tau}_{\nu \sigma}-\Gamma^{\mu}_{\tau \sigma} \Gamma^{\tau}_{\nu \rho},\\
	T^{\mu}_{\nu \rho} = \Gamma^{\mu}_{\rho \nu}-\Gamma^{\mu}_{\nu \rho},\\
    Q_{\mu \nu \rho} = \nabla_{\mu} g_{\nu \rho} = \partial_{\mu} g_{\nu \rho} - \Gamma^{\lambda}_{\mu \nu} g_{\mu \rho} - \Gamma^{\lambda}_{\mu \rho} g_{\lambda \nu}.
\end{align}
As one can see in Fig. \ref{fig:Fig1}, the curvature tensor encodes the variation of the angles in a parallel transport along a closed curve on a manifold; the torsion tensor encodes how the tangent space twists around a curve when we parallel transport two vectors along each other; non-metricity encodes the variation of vectors’ length when they are moved along a curve \citep{bahamonde2023teleparallel}.

\begin{figure}
    \centering
    \includegraphics[height=8.95cm]{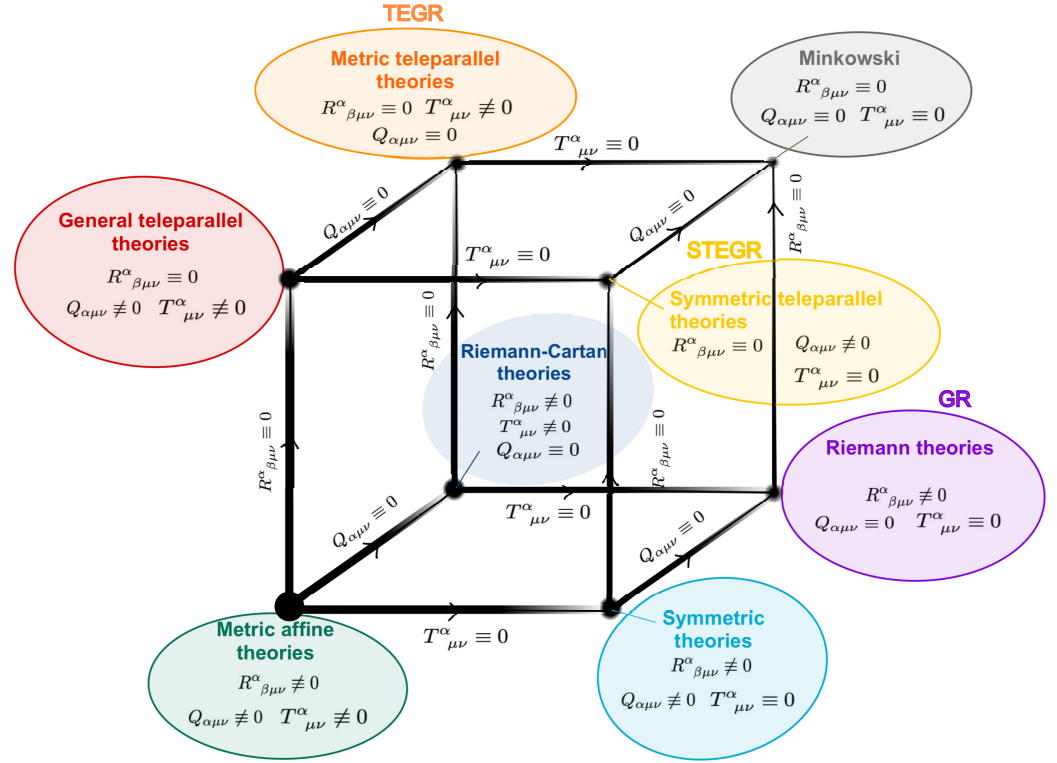}
    \caption{ A map of the possible metric-affine theories. GR is a particular theory where torsion and non-metricity are set to zero (see also \citep{bahamonde2023teleparallel,capozziello2022comparing}). }
    \label{fig:Fig3}
\end{figure}

With these three geometrical objects, we can build all the possible metric-affine theories, as can be seen in Fig. \ref{fig:Fig3}.

As anticipated, several authors \citep{capozziello2022comparing, capozziello2024equivalence, krvsvsak2019teleparallel, beltran2019geometrical} claim that TEGR and STEGR can be formulated  to be equivalent to GR in multiple features. Thus, in the following we are going to analyse the features in which the equivalence arises.

\subsection{Equivalence of Lagrangians} \label{Equivalence of Lagrangians}
Firstly, there is an equivalence at the Lagrangian level. In fact, GR dynamics can be derived from the Hilbert-Einstein
action \citep{capozziello2022comparing}:
\begin{align} \label{eq:HE_action2}
	S_{GR} = \frac{c^4}{16 \pi G} \int  d^4x \sqrt{-g}~ (\mathcal{L}_{GR} + \mathcal{L}_{m}), 
\end{align}
where $\mathcal{L}_{GR}=\mathcal{R}$ and $\mathcal{R}$ is the Ricci curvature scalar;  $\mathcal{L}_m$ is the matter Lagrangian. In TEGR and STEGR the same dynamics can be recovered, up to a boundary term. 

Teleparallel gravity can be built in different ways (see \citep{March2025equivalence} for a recent review). A useful approach to formulating it as a gauge theory \citep[p. 16]{March2025equivalence} is to work with tetrads $e^A_{\mu}$, which describe gravity, and spin connections $\omega^A_{B\mu}$, which account for inertial effects. We follow this tradition \citep{krvsvsak2019teleparallel, capozziello2022comparing, aldrovandi2012teleparallel}, not least because it provides an elegant formulation of the SEP. Tetrad fields are geometric constructions which establish a relation between the manifold and its tangent spaces as a soldering agent. On the other hand, the spin connections account for inertial effects in rotated frames. The coordinates of the two frames are related with Lorentz transformations.

Thus, in TEGR, with $\mathcal{L}_{TEGR}=-T$, i.e. the torsion scalar, we have \citep{capozziello2022comparing}:
\begin{align} \label{eq:TEGR_action}
	S_{TEGR} = \frac{c^4}{16 \pi G} \int  d^4x~ e~ \mathcal{L}_{TEGR} + \int d^4x~ e~ \mathcal{L}_{m} \\
	\mathcal{R}=-T-\frac{2}{e} \partial_{\mu} (eT^{\mu})\\
	T=\frac{1}{2}S_A^{\mu\nu}T^A_{\mu\nu}= \frac{1}{2}(K_A^{\mu\nu}-e_A^{\nu}T^{\mu}+e_A^{\mu}T^{\nu}) T^A_{\mu\nu},
\end{align}
where $T^{\alpha\mu}_{\alpha}=T^{\mu}$ is the torsion vector and $K^C_{BA}$ the contortion tensor. $S_A^{\mu\nu}$ is the superpotential $ S_A^{\mu\nu}=K_A^{\mu\nu}-e_A^{\nu}T^{\mu}+e_A^{\mu}T^{\nu}$. And $e$ denotes the determinant of $e_A^{\mu}$.

Similarly, in STEGR, with $\mathcal{L}_{STEGR}=Q$, the non-metricity scalar, we have \citep{PhysRevD.105.024042, beltran2019geometrical}:
\begin{align} \label{eq:STEGR_action}
	S_{STEGR} = \frac{c^4}{16 \pi G} \int  d^4x~ \sqrt{-g}(\mathcal{L}_{STEGR} +  \mathcal{L}_{m})\\
	Q=g^{\mu \nu} (L^{\alpha}_{\beta \mu}L^{\beta}_{\nu \alpha}-L^{\alpha}_{\beta \alpha}L^{\beta}_{\mu \nu})= \mathcal{R}+\nabla_{\mu} (Q^{\mu}-\bar{Q}^{\mu}),
\end{align}
where $Q_{\alpha}=Q_{\alpha \lambda}^{\lambda}$ and $\bar{Q}_{\alpha}=Q^{\lambda}_{\alpha \lambda}$.

First of all, the equivalence among the three theories is evident at Lagrangian level. In fact, the actions of the three theories differ by boundary terms, which are dynamically passive, since the spacetime is asymptotically flat at infinity and thus these terms give no contribution \citep{capozziello2022comparing}. As mentioned, this equivalence does not hold for extensions like $f(\mathcal{R})$, $f(T)$, and $f(Q)$ \citep{Capozziello2023boundary}.

\subsection{Equivalence of the field equations}
Secondly, the same comparison can be developed at the level of field equations. We can start from the Bianchi identities, which have the important role to link the field equations with the conservation laws of the gravity tensor invariants and the energy-momentum tensor. Also in this case,   the equivalence of the three formulations  can be achieved.

The most general second Bianchi identity is the following \citep{bahamonde2023teleparallel}:
\begin{align} \label{eq:bianchi}
	\nabla_{\lambda}R^{\alpha}_{\beta\mu\nu}+\nabla_{\mu}R^{\alpha}_{\beta\nu\mu}+\nabla_{\nu}R^{\alpha}_{\beta\lambda\mu}= T^{\rho}_{\mu\lambda}R^{\alpha}_{\beta\nu\rho}+ T^{\rho}_{\nu\lambda}R^{\alpha}_{\beta\mu\rho}+ T^{\rho}_{\nu\mu}R^{\alpha}_{\beta\lambda\rho}\,.
\end{align}
In GR, since we have no torsion and non-metricity, we derive the Einstein field equations (EFE) in vacuum:
\begin{align} \label{eq:GR_fe}
	\nabla_{\mu}(\overset{\circ}{R^{\mu\nu}}-\frac{1}{2}g^{\mu \nu}\overset{\circ}{R})=0\,, 
\end{align}
where the notation $\overset{\circ}{R}$ stands for quantities built up on the Levi-Civita connection, i.e. in this case the Ricci tensor.

In TEGR, having vanishing curvature and non-metricity, via the Weitzenböck gauge, we obtain an equivalent expression of the EFE which is \citep{capozziello2022comparing}:
\begin{align} \label{eq:TEGR_fe}
    R^{\alpha}_{\beta\mu\nu}=\overset{\circ}{R^{\alpha}}_{\beta\mu\nu}+K^{\alpha}_{\beta\mu\nu}\\
	R_{\mu\nu}-\frac{1}{2}g_{\mu \nu}R=-K_{\mu\nu}+\frac{1}{2}g_{\mu \nu}K\\
    K_{\mu\nu}-\frac{1}{2}g_{\mu \nu}K=0
\end{align}
Similarly, in STEGR, since there is no curvature and torsion, and via the coincident gauge, we obtain equivalent field equations \citep{capozziello2022comparing}:
\begin{align} \label{eq:STEGR_fe}
	R^{\alpha}_{\beta\mu\nu}=\overset{\circ}{R^{\alpha}}_{\beta\mu\nu}+L^{\alpha}_{\beta\mu\nu}\\
    R_{\mu\nu}-\frac{1}{2}g_{\mu \nu}R=-L_{\mu\nu}+\frac{1}{2}g_{\mu \nu}L\\
    L_{\mu\nu}-\frac{1}{2}g_{\mu \nu}L=0 
\end{align}
Therefore, despite being built on different principles and geometric objects, equivalent formulations of the EFE can also be derived in TEGR and STEGR.

Since the same field equations have been obtained, the same exact solutions, under the same symmetries and boundary conditions, have to be achieved. The Schwarzschild solution, i.e. the spherically symmetric solution, and the validity of the Birkhoff theorem can be derived in all the three formulations \citep{capozziello2022comparing, krvsvsak2019teleparallel}. This is important, since if the field equations and their solutions are the same, we have the same empirical predictions. For instance, the classic tests of the trajectories of massive bodies and photons, according to the Schwarzschild solution of the field equations, confirm in the same way GR, TEGR and STEGR. So we cannot say anymore that these classic tests are corroborations of GR and of spacetime curvature  \citep{wolf2023underdetermination}.

\subsection{The role of  the Strong Equivalence Principle}
Finally, theorists have found another significant empirical equivalence, that is the recovery of the SEP  \citep{krvsvsak2019teleparallel, capozziello2022comparing}.

As for TEGR, one can also formulate GR in the tetrad formalism, where the Levi-Civita connection is translated into a Lorentz connection \citep{beltran2019geometrical, capozziello2022comparing}:
\begin{align} \label{eq:tetrads_GR}
	\overset{\circ}{\omega}^A_{B \mu}= e^A_{\lambda}e^{\nu}_B\Gamma^{\lambda}_{\mu \nu}+e^A_{\sigma}\partial_{\mu}e^{\sigma}_B=e^A_{\nu}\nabla_{\mu}e^{\nu}_B,
\end{align}
where $\overset{\circ}{\omega}^A_{B \mu}$ accounts for both gravitational and inertial effects in GR.  From the General Covariance Principle in its active formulation, we can obtain the coupling gravitational prescription. This principle states that the laws of physics - initially written in a special-relativistic context - must retain their form also in the presence of gravity, requiring invariance under spacetime diffeomorphisms. In the case of GR, for a general field $\Psi$ \citep{krvsvsak2019teleparallel, capozziello2022comparing}, it is:
\begin{equation} \label{eq:GR_prescription}
\partial_\mu \Psi \quad \longrightarrow D_\mu \Psi = \partial_\mu \Psi + \frac{1}{2} \overset{\circ}{\omega}{}^C_{\ B\mu} S^B_{\ C} \Psi
\end{equation}
where $S^B_{\ C}$ are the Lorentz generators.

The analogous coupling in TEGR can be similarly derived. Transitioning from the special-relativistic to the case with gravity, incorporating the appropriate contortion tensor in the Lorentz covariant derivative, we obtain:
\begin{align}\label{eq:TEGR_prescription}
\partial_\mu \Psi \quad \longrightarrow D_\mu \Psi = \partial_\mu \Psi + \frac{1}{2} \left( \omega^C_{\ B\mu} - K^C_{\ B\mu} \right) S^B_{\ C} \Psi
\end{align}
The combination of (\ref{eq:GR_prescription}) and (\ref{eq:TEGR_prescription}) yields the relation between the gravitational and inertial effects in TEGR and GR  \citep{krvsvsak2019teleparallel, capozziello2022comparing}:
\begin{align} \label{eq:TEGR_GR_relation}
	\omega^C_{B \mu}-K^C_{B\mu}=\overset{\circ}{\omega}^C_{B \mu},
\end{align}
where $\omega^C_{B\mu}$ and $K^C_{B\mu}$ account respectively for inertia and gravitation in TEGR, and ${\omega}^C_{B \mu}$ for both gravitation and inertia in GR. Therefore, choosing a LIF,  where the GR spin connection vanishes $\overset{\circ}{\omega}^A_{B \mu}=0$, we obtain the identity between the inertial effects and gravitation in TEGR \citep{krvsvsak2019teleparallel, capozziello2022comparing}: 
\begin{align} \label{eq:TEGR_SEP}
	\omega^C_{B \mu}=K^C_{B\mu},
\end{align}
which recover the second formulation of  SEP defined in Sec. \ref{General Relativity assumptions and difficulties}, SEP\(_{2,\mathrm{\neg EEP}} \), which does not assume the EEP, or in the words by Read and Teh, it does not conceptually unify gravity and inertia \citep[p. 3486]{Read2023}. In fact, here it is evident how in TEGR we have the separation of gravitational and inertial effects, the former identified by the contortion tensor and the latter by the spin connection. For some authors, this possibility of separation is one of the most important properties of TEGR \citep{krvsvsak2019teleparallel}. 

For a recovery  of SEP\(_{1,\mathrm{ EEP}} \) in the TEGR context, see \citep[p. 3491]{Read2023}.\\

With respect to STEGR,  writing its connection by tetrads $e^{\alpha}_{\beta}$ \citep{capozziello2022comparing}:
\begin{align} \label{eq:STEGRt_gauge}
	\Gamma^{\alpha}_{\mu\nu}=(e^{-1})^{\alpha}_{\beta}\partial_{\mu}e^{\beta}_{\nu}\,,
\end{align}
without curvature and torsion, and by a particular transformation of coordinates in a point $p$, we can arrive at the coincident gauge \citep{PhysRevD.105.024042, capozziello2022comparing, read2025clarifying}:
\begin{align} \label{eq:coincident_gauge}
	\Gamma^{\alpha}_{\mu\nu} (p)=\frac{\partial x^{\alpha}}{\partial \xi^{\lambda}}\partial_{\mu}\partial_{\nu}\xi^{\lambda}=0\,. 
\end{align}
The vanishing of the connection physically means that the origin of the tangent flat space and the one of the manifold are coincident \citep{capozziello2022comparing}, which is the formulation of the SEP defined in Sec. \ref{General Relativity assumptions and difficulties}. The SEP is then recovered also in STEGR, via the coincident gauge. 

Recovery of been considered one of the most important ways in which TEGR and STEGR demonstrate their empirical equivalence with GR \citep{capozziello2022comparing, krvsvsak2019teleparallel}. Note also that this empirical equivalence is distinct from the equivalence given by the dynamics, i.e. by the action (Sec. \ref{Equivalence of Lagrangians}). In fact, as Wolf $\&$ Read show, the empirical content of a theory is not exhausted by its dynamics \citep{Wolf_2023boundaries}.
The recovery of the SEP in the explored regimes and at the tested levels is, of course, essential for any theory of gravity to be considered consistent, given the amount of high-precision experimental tests (see Sec. \ref{Experimental perspectives} for the state of the art). Therefore, these results are highly significant. However, in TEGR and STEGR, the SEP is not postulated as fundamental, as it is in GR, but instead emerges as the result of a possible gauge choice. This structural difference, as we will argue in the following sections, could imply differences in empirical content.\\


In conclusion of this section, we can summarize the significant results in  Geometric Trinity as: 
\begin{itemize}  
\item The equivalence  at  Lagrangian level (up to a boundary term) holds.
\item The equivalence of  field equations holds  starting from the general second Bianchi identities.
\item The same solutions of the field equations are recovered in all the three theories.
\item The SEP is recovered in TEGR and STEGR, even though such a principle is not at their foundation.
\end{itemize}
In the following section, we are going to discuss    these equivalences with particular focus on their epistemic and experimental implications.

\section{Epistemological considerations on Equivalent Gravities}\label{Epistemological considerations on Equivalent Gravities}
\subsection{High degree of under-determination}\label{High degree of under-determination}
We have seen that there is a high degree of under-determination among the Geometric Trinity of Gravity, as already pointed out both in the physical \citep{capozziello2022comparing, krvsvsak2019teleparallel, beltran2019geometrical} and the foundational \citep{mulder2023spacetime, march2023geometric, wolf2023underdetermination, knox2011newton} literature. This causes both various epistemological and metaphysical problems. In this work we focus particularly on \textit{if} and \textit{how} we can distinguish experimentally among them. Although the three theories are dynamically equivalent, and the recovery of  SEP is often regarded as one of the key features through which TEGR and STEGR demonstrate their empirical equivalence with GR, we argue that the differing foundational status of  SEP across the three theories may nonetheless lead to differences in empirical content.  We think that this difference is strictly related to the assumptions of GR on the spacetime structure and on the fundamental objects of the theory. 

However, before coming into the details of the central argument of this work, it is important to introduce the epistemic tools that will be exploited. They  are extensively discussed among philosophers of physics and epistemologists in the  context of high degrees of under-determination.

The landscape of theoretical physics has evolved in the last 2-3 decades, because for most theories beyond the Standard Model of Particles and  Quantum Gravity, empirical data are either scarce or completely absent. Nevertheless,theories like  ST, Supersymmetry or Cosmic Inflation have all been defended for decades, although none of the classical methodologies seem to straightforwardly apply. While experimental testing remains the gold standard, the use of analogue experiments and the so-called non-empirical ways of theory assessment, or meta-empirical, have been proposed, especially in fundamental physics and cosmology. The issue is particularly relevant because today, fundamental physics clearly is not driven by perspectives of technological utilization in a few years, and the typical time scale for that intermediate state has grown beyond one generation of scientists. During most of the 20th century, fundamental physics was perceived as a scientific field where theories typically could be empirically tested within a reasonable time frame. But today the situation is different. 
Moreover, even concerning proper experimental tests, contemporary experiments are far more intricate, and the evaluation and interpretation of the data are subtle and by no means trivial matters. 

These are some of the reasons why some philosophers of physics have delved into this idea and formulated theories of \lq\lq confirmation" that make the corresponding intuition more rigorous. This approach (see for instance \citep{dawid2013string, dawid2022meta, dawid2019significance}) exploits Bayesian Confirmation Theory and deeply relies on the practice of exploring and constraining the theory space. In fact, when scientists find that despite substantial efforts, no alternative viable hypothesis are capable of explaining some scientific problem, they tend to place more trust in the existing theory. This is indeed called the \lq\lq No Alternatives Argument". These approaches are called \lq\lq meta-empirical", since they are not empirical in the common sense, but still involve  observations. For a general Bayesian formalization of this argument and a proof that it counts as evidence, see Ref. \cite{dawid2015no}, as well as for its limits.

Note that we have not to confuse the use of the concept of non-empirical \lq\lq confirmation" as the confirmation of a theory in the traditional sense. 


Anyway, in this framework of constraining theory space, authors are developing some interesting tools to address cases of under-determination and assess untested hypotheses among competing theories. 

In Ref. \citep{dardashti2019physics}, the author develops an epistemological reflection on the theoretical exploration of alternative theories, which fits perfectly with the material of this work, allowing for a more precise formulation of the argument. In the following, the general argument of \citep{dardashti2019physics} will be briefly introduced, and then it will be applied to our specific case. See Ref. \citep{dardashti2019physics} also for the limits of the approach.\\

Let us assume we are interested in whether we can trust the predictions of some theory $Th$. We have made a large set of observations which are in agreement with the prediction $P_1$ of $Th$ and therefore confirm it. Suppose that $Th$ also makes the predictions $P_2$ and $P_3$. We usually will have some confidence in these predictions of $Th$, as it has so far been an empirically successful theory. So the previous empirical success warrants an increase in our trust regarding the novel predictions $P_2$ and $P_3$ of $Th$. 

Now let's assume that for some reason we will not be able to conduct experiments on $P_2$ and $P_3$. In these circumstances, we cannot further assess these predictions based on empirical data. Now assume  that someone comes up with an alternative theory, say $Th'$, which happens to also predict the set of observations $P_1$ and it is therefore similarly confirmed by it. In addition, $Th'$ predicts $P_2$ but disagrees about $P_3$. Let us denote the predictions by:

\begin{quote}
   \large{ Predictions($Th$) = \{$P_1$, $P_2$, $P_3$...\}\\
    Predictions($Th'$) = \{$P_1$, $P_2$, $\neg P_3$ ...\}}
\end{quote}
How will the existence of this additional theory impact ones believe regarding the predictions $P_2$ and $P_3$? The same available empirical data, i.e. $P_1$, confirms two competing theories, which agree with respect to one prediction, $P_2$, and disagree with respect to another prediction, $P_3$. If we have no reason to trust one theory more than the other, then the proposal of the competing theory $Th'$ should lead to an increase in our trust regarding the prediction $P_2$, while it leads to a decrease with respect to the prediction $P_3$. Now imagine further, scientists come up with another theory $Th''$, which agrees with respect to the prediction $P_2$ and disagree with respect to $P_3$:

\begin{quote}
    \large{Predictions($Th$) = \{$P_1$, $P_2$, $P_3$...\}\\
    Predictions($Th'$) = \{$P_1$, $P_2$, $\neg P_3$ ...\}\\
    Predictions($Th''$) = \{$P_1$, $P_2$, $\neg P_3$ ...\}}
\end{quote}
It is reasonable to assume that we would slowly become more and more certain about $P_2$ being a feature of the world we live in but not about $P_3$. This is more evident when we have agreement on multiple non-trivial risky predictions:
\begin{quote}
    \large{Predictions($Th$) = \{$P_1$, $P_2$, $P_3$, $P_4$, $P_5$, ...\}\\
    Predictions($Th'$) = \{$P_1$, $P_2$, $P_3$, $P_4$, $\neg P_5$, ...\}\\
    Predictions($Th''$) = \{$P_1$, $P_2$, $P_3$, $P_4$, $\neg P_5$, ...\}}
\end{quote}
Therefore, if this argument is right, this counts as an evidence, a meta-empirical evidence, against the hypothesis $P_5$. This counts not as an $empirical$ observation, but as a $meta$-empirical observation. Counting as an observation, it affects the posterior probability of the validity of $P_5$, as other empirical evidence \citep{dawid2015no}. Again, if also all the other competing theories would have had $P_5$ as a prediction, this would have been counted as a meta-empirical evidence in its favor. To be clear, in no way this is a posterior evidence with the same strength of an empirical one, but it still counts as evidence.

In this way, the exploration of competing alternatives allows us to better assess the untested predictions of the theory. Therefore, the practice of exploring theory space is highly powerful especially in contexts with high degrees of under-determination \citep{dardashti2019physics}. As anticipated, the context of  Geometric Trinity is properly one of them. In fact, one of the main problems of this debate is that, even conceptually, it turns out that it is difficult to sharply distinguish between predictions of different theories of gravity. This means that they often do not lead to clear observational differences. The result is that every proposal that is viable  mimics every other proposal that is viable, both empirically and conceptually. This is primarily because modified gravity scenarios, both extensions and alternatives, are victims of the GR success, so they have to reproduce its phenomenology in many features.

\subsection{The debate  on the Equivalence Principle}

As we have seen, despite the fact that TEGR and STEGR are built on  different foundation principles with respect to GR, physicists claim that they both  recover the SEP, which is of course a necessary condition for a consistent theory of gravity, at least at classical level. However,  it is often overlooked that there is a crucial difference in their relation with the SEP. In  GR,  SEP is a fundamental assumption of the theory, while, in TEGR and STEGR, it is not postulated \textit{a priori}  but it is recovered \textit{a posteriori}. As it will be argued here, this structural difference could represent a possible important distinction in the empirical content of theories in Geometric Trinity, since if it is not fundamental, there is the possibility that, at some level, it could be not valid. 

Moreover, this result seems sufficient to regard TEGR and STEGR as different proper theories, instead of mere mathematical reformulations of GR, as some authors have suggested based on their dynamical equivalence \citep{knox2011newton}. According to that perspective, it was  possible that the under-determination among the Geometric Trinity would later turn out to be an ill-posed problem. In contrast, recent theoretical advancements seem to have clarified this point. This is consistent with other works that acknowledge the lack of categorical equivalence between GR and TEGR \citep{weatherall2024general, weatherall2025some, read2025clarifying, March2025equivalence}.

Given  the results highlighted in Sec. \ref{Equivalent Gravities}, we can now apply the epistemological consideration just introduced above, that is:
\begin{multicols}{2}
\begin{quote}
\small{
 Pred.($Th$)~ = \{$P_1$, $P_2$, $P_3$, $P_4$, $P_5$\}\\
Pred.($Th'$) = \{$P_1$, $P_2$, $P_3$, $P_4$, $\neg P_5$\}\\
Pred.($Th''$) = \{$P_1$, $P_2$, $P_3$, $P_4$, $\neg P_5$\}\\
     Pred.($GR$)~~~~~~~= \{$L$, $FE$, $S$, $C$, $FEP$\}\\
     Pred.($TEGR$) ~~= \{$L$, $FE$, $S$, $C$, $\neg FEP$\}\\
     Pred.($STEGR$) = \{$L$, $FE$, $S$, $C$, $\neg FEP$\}
     }
\end{quote}
\end{multicols}
On the left, the general epistemic argument  is shown given by \citep{dardashti2019physics} and explained before; on the right there is the application to Trinity Gravity. It seems that this case  fits perfectly with the above general epistemological considerations.

Following the five predictions discussed in \ref{Equivalent Gravities}, $L$ stands for the equivalence at the Lagrangian level, $FE$ for the field equations derived from the second Bianchi identity, and $S$ for the solutions of the $FE$. Then $C$ stands for cosmological applications.  In fact, cosmological observations can be considered very important evidence for GR, but since we can now build cosmological models also with TEGR and STEGR, their predictive power in cosmology should be taken into account as well. In fact, people are already studying cosmological applications of TEGR and STEGR (see for instance \citep{cai2016f, heisenberg2024review}). So we cannot anymore say that cosmological observations are evidence for GR only. 

Finally, $FEP$ stands for  \textit{Fundamental Equivalence Principle}. As demonstrated before, in TEGR and STEGR, the  SEP is not fundamental, and so we can write $\neg FEP$. In other words, we can say that TEGR and STEGR predict an \textit{EMergent Equivalence Principle}, or $EMEP$, instead of a $FEP$. This does not mean a prediction for a violation of the  SEP, but only that TEGR and STEGR do not have the  SEP at their foundation, i.e. they do not share the prediction $FEP$. In fact, the prediction $EMEP$ is, in some sense, shared by all the three theories, since an $EMEP$ is contained in $FEP$. If the  SEP is fundamental, it has to be always valid at any level, and so it is also valid at emergent levels, but the contrary is not necessarily the case. In other words, the $FEP$ implies the $EMEP$, but the $EMEP$ does not imply the $FEP$:
\begin{align} \label{eq:EMEP}
FEP \rightarrow EMEP\\
EMEP \not\rightarrow FEP
\end{align}
The meaning of this difference can be seen more clearly in the geometric definition of  SEP. As observed, in GR, where the $FEP$ holds, it must always be possible to find a LIF in which gravitational effects can be nullified. In contrast, in TEGR and STEGR, this is not necessary, although it remains possible.

So, there is now a new beast in the bunch of beasts  \citep{lehmkuhl2021equivalence}, the $EMEP$.

\begin{figure}
    \centering
    \includegraphics[height=4.8cm]{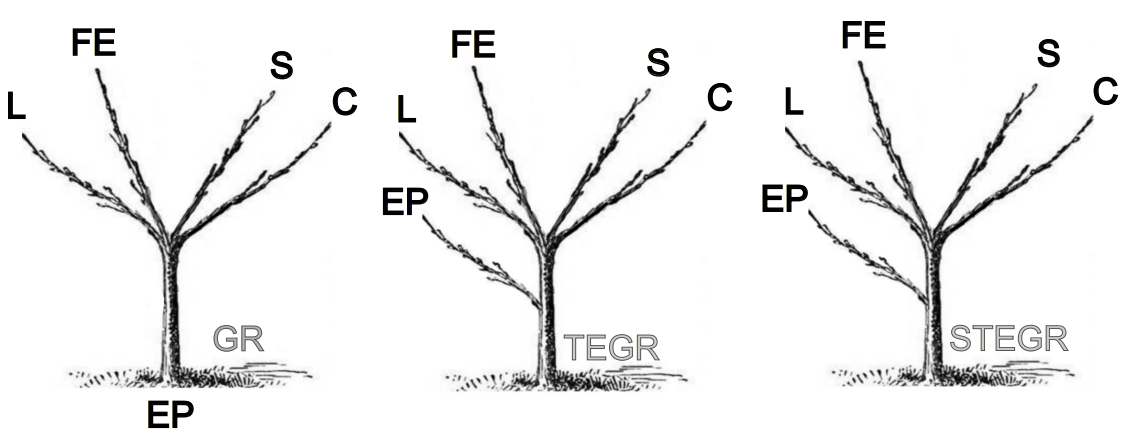}
    \caption{The figure shows how the three theories differ with respect to the EP (the SEP, in particular). In GR, EP is at the foundation of the theory and without it the other predictions are not possible; without the root, the branches would not exist.  In TEGR and STEGR, on the contrary, the EP is not at the foundation, but it is a lateral branch, recoverable  through the general covariance principle and via the coincident gauge, respectively.
    $L$: Lagrangians, $FE$: field equations, $S$: solutions of the $FE$, $C$:  cosmological applications.}
    \label{fig:Fig2}
\end{figure}
As argued, this situation decreases our confidence in that hypothesis on which the equivalent theories do not agree. That is to say, if the argument is correct, our confidence in the fundamentality of the  SEP is decreased.

This argument is also independent of the history of the theory itself. In fact, imagine that both TEGR and STEGR would be built only with the  SEP as a foundation; imagine that $L$, $FE$, $S$ and $C$ were derivable only if $FEP$ was valid, i.e. only assuming the  SEP. Then our confidence in the  SEP as a necessary principle of any consistent theory of gravity would be increased. In the terms previously defined, this observation would have counted as a meta-empirical evidence for the hypothesis $FEP$, and consequently for a metric theory of gravity. It would have been considered evidence in favor of $FEP$ because, after searching for an alternative hypothesis to $FEP$ to build a viable and coherent theory of gravity, physicists would not have found it. Therefore, as shown above, the posterior probability of the $FEP$ hypothesis would be increased. But the results are pointing to exactly the opposite. In fact, both TEGR and STEGR can recover the  SEP, but it is not at their foundation, i.e. they do not predict $FEP$. See Fig. \ref{fig:Fig2}.

More precisely, any viable theory of gravity has to recover the  SEP, at least at the scales and at the levels of accuracy of the present experiments. So the fact that TEGR and STEGR can recover the  SEP is an important feature. This is one of the reason why it is  difficult to distinguish empirically among the three theories, since TEGR and STEGR  necessarily have to recover  the experimental tests of GR. But properly the fact that the  SEP is not at the foundation of them constitutes a possible important difference from the empirical content.

In other words, this argument shows that   SEP is not a principle necessary present in all  viable theories of gravity, but it is an assumption on which GR is built. 

\section{Discussion}\label{Discussion}
\subsection{Implications and limits of the argument}\label{Implications and limits of the argument}
In the general example presented at the beginning of Sec. \ref{Epistemological considerations on Equivalent Gravities}, we mentioned  non-tested hypotheses. In this perspective, one could think that the  SEP is instead a well tested hypothesis. 
However, the argument does not address the SEP itself, but rather the \textit{fundamentality} of  SEP, which has not been decisively tested — as other authors have suggested, questioning the \textit{universality} of  EP rather than its \textit{fundamentality}, using similar arguments based on teleparallel gravity \citep[ch. 9]{aldrovandi2012teleparallel}.  In fact, particularly at quantum level, the WEP, which is at the foundation of  SEP, is tested but not decisively tested \cite{Torrieri:2024ivy}. The situation would instead radically change if  a  fundamental theory, requiring  some form of  EP at any level, were formulated.

On the other hand, one could  include, in the argument, all  theories  predicting the violation of the  SEP,  in order to show that also  these theories disagree with the hypothesis $FEP$. However, these theories are not really dynamically equivalent to GR as in the case of TEGR and STEGR. Even if some of them would  recover  the GR phenomenology, given the state of the art today,  they  cannot be considered the final theory of gravity. They may have good empirical and observational 
evidences in different regimes and scales, as in the case of MOND, but they give not the same predictions 
$P_1,...,P_n$. Therefore, the same epistemic considerations would not be valid.\\

Then, we can briefly see how the eight conceptual difficulties presented in Sec. \ref{General Relativity assumptions and difficulties} are, at least, mitigated by these results.

The first two problems $(i)$ and $(ii)$ can be addressed thanks to the Palatini formalism, where the metric $g_{\mu\nu}$ and the connection $\Gamma^{\rho}_{\mu \nu}$ are independent, with $g_{\mu\nu}$ determining the causal structure and $\Gamma^{\rho}_{\mu \nu}$ the free fall, i.e. the geodesic structure. In this way, the connection becomes the true fundamental dynamical variable and the observable of the theory. Following the words by Einstein,  the metric is \lq\lq dethroned" and becomes an \lq\lq ancillary variable" (see Ref.  \cite{ferraris1982variational} and references therein).

The tension with  new physics predictions $(iii)$, with quantum mechanics $(iv)$ and the unjustified {\it a priori } coincidence between $m_I$ and $m_G$ $(vii)$ would be resolved, since the  SEP, along with the WEP and the EEP, could be considered an emergent feature, recoverable but not postulated. So the only requirements would be the compatibility of  theories with experimental constraints. Then, TEGR and STEGR allow gravity to be reformulated as a gauge theory $(vi)$, and their extensions $f(T)$ and $f(Q)$ could be studied in order to search for a geometrical solution to DE and CDM problems $(v)$.

Finally, without imposing the  SEP, we are not privileging the spacetime structure  over torsion or non-metricity \textit{a priori} (viii). The geometric under-determination would remain, but the correct spacetime structure would no longer be determined by a postulate.\\

With respect to the limits of the argument, there are a couple of useful considerations. First, the argument would not be valid if TEGR and STEGR were revealed as inconsistent theories, since we would no longer trust their predictions. A similar predictive power of the competing theories is a necessary condition for the soundness of the argument. So if our trust in TEGR and STEGR would decrease for some reason, the argument would be not valid anymore.

Second, it could be objected that the number of true predictions which would be considered sufficient in order to trust the further predictions of the theories is somewhat arbitrary. Are $four$ good predictions sufficient? Actually, in this case, there are five good predictions, since the $EMEP$ is a true non-trivial and very significant prediction of the theories. One could point out that predictions $S$ and $C$ are actually consequences of prediction $FE$, since both the solutions and the cosmological applications depend on the field equations. This is true; maybe they are not independent predictions, but the interesting thing is again that with the extensions $f(T)$ and $f(Q)$, the equivalence is not anymore valid \cite{Capozziello2023boundary}, so it is important to explore any possible different feature in solutions and applications \citep{cai2016f, heisenberg2024review}.

\subsection{Experimental perspectives}\label{Experimental perspectives}
As mentioned, physicists are not yet satisfied with the current precision of the tests of the different formulations of the EP, as we can see from the numerous experiments which are  developed by many different research groups \citep{tino2020precision}. These tests are complex and require many years of work and experimental efforts, not to mention the proposed space missions. All this efforts are developed in order to increase the accuracy of the EP tests, demonstrating the unsatisfactory situation, especially at quantum level.

Our argument does not mean  that  SEP is false or that it has to be necessarily violated at some level, but surely it encourages experimentalists in the search for a possible violation  to discriminate among concurring theories of gravity. If  the  SEP is not fundamental, it could be an emergent property. Therefore, it could be the case that it is violated at some level, for example at quantum level. 
The incoming experiments of free falling with quantum tests could be the straightforward approach to probe the above statement (see for instance \citep{Rosi_2017, tino2020precision, Tino_2021}). Clearly these quantum tests are WEP tests, and $\neg$SEP does not imply $\neg$WEP, but $\neg$WEP does imply $\neg$SEP, i.e. if we detect a violation of the WEP, this would falsify also the first formulation of  SEP given in Sec. \ref{General Relativity assumptions and difficulties}.  This is relevant because the finest experiments on  EP we have at the moment are  conceived for the weak formulation. Clearly, the free fall of a wave packet is something different with respect to the free fall of a classical test particle. This conceptual aspect needs further and deep investigations.

Currently, the highest accuracy on the Eötvös parameter, which quantify the violation of the WEP, has been reached by the MICROSCOPE space mission with a free-fall experiment performed with
macroscopic classical masses. In 2017, they reached $10^{-14}$ \citep{touboul2017microscope} and, in 2022, $10^{-15}$ \citep{touboul2022m}. Other future space missions have been proposed, such as the Galileo Galilei (GG) \citep{nobili2018testing} and the Satellite Test of the Equivalence Principle (STEP) \citep{overduin2012step}, with the goal of achieving $10^{-17}$ and $10^{-18}$, respectively. The atomic experiments have compared the free fall of different isotopes or atomic species such as $^{85}$Rb and $^{87}$Rb, $^{39}$K and $^{87}$Rb, the bosonic $^{88}$Sr and the fermionic $^{87}$Sr and also atoms in different spin orientations. In an experiment in Stanford \citep{asenbaum2020atom}, a precision of $10^{-12}$ was reached 
(see \citep{tino2020precision,Tino_2021} for a review).  
Similarly to the classical counterparts, the ultimate performance of atomic sensors for WEP tests can be reached in space, where tests with a precision of $10^{-15}\div 10^{-17}$ were proposed by the STE-QUEST (Space–Time Explorer and QUantum Equivalence Space Test) mission \citep{altschul2015quantum, struckmann2024platform}. Experiments exploiting entangled atomic states aim to push the sensitivity beyond the so called Standard Quantum Limit \citep{Salvi_2018}.

Finally, although the technology is not yet mature for high-precision measurements, there has been progress in developing experiments to test the WEP with antimatter \citep{tino2020precision, Vinelli_2023, Anderson2023}. This line of research is particularly interesting, as a WEP test using a matter-wave interferometry with antihydrogen or positronium would probe both the quantum and antimatter regimes, domains in which the validity of the WEP remains uncertain.

\begin{figure}
    \centering
    \includegraphics[height=7.8cm]{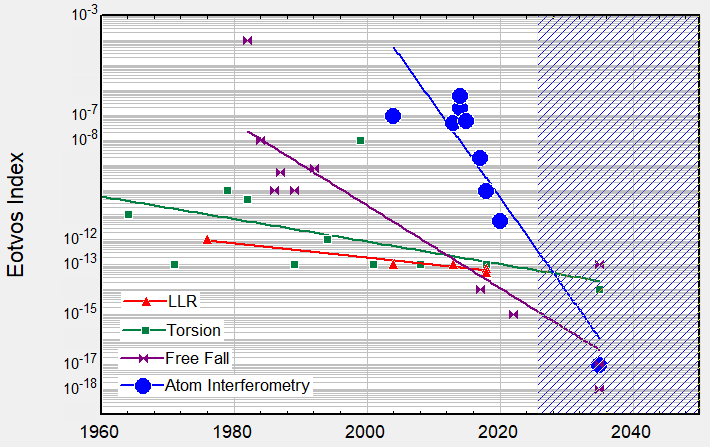}
    \caption{
    The most relevant WEP tests performed from 1960. Quantum WEP tests with atom interferometry are not yet at the same accuracy as their classical counterparts, but it is a novel technology (20-30 years) and its development is impressively faster than the classical tests. In the shaded area on the right there are the prospects of future missions and projects (for the data, see \citep{asenbaum2020atom, altschul2015quantum, struckmann2024platform, tino2020precision, nobili2018testing, touboul2022m}).}
    \label{fig:Fig4}
\end{figure}

Fig. \ref{fig:Fig4} shows the limits set by the WEP tests performed with different methods, from 1960 until today and the future prospects. For the   SEP tests in the sense of Will \citep[p. 76]{Will2018theory}, the achieved accuracy limit is $\eta_{SEP} \approx 10^{-5}$ \citep{tino2020precision}.

If a violation of some form of the EP were to be detected, it would constitute a falsification of GR at that level, while TEGR and STEGR would remain viable as theories. This point is controversial. Some authors argue that, since the three nodes of the geometric trinity are dynamically equivalent, no empirical differences between them should exist \citep{weatherall2025some}. According to this view, a violation of the SEP would falsify not only GR, but also TEGR and STEGR at that level. However, GR postulates that the SEP is fundamental and holds at every point in spacetime, whereas in TEGR and STEGR it can be recovered through possible gauge choices. For them, in principle there may exist points where the SEP does not hold. If experiments were to reveal that the SEP fails at some point, the gauge choices that restore the SEP at those regions would constitute surplus structure, unnecessary for representational purposes \citep{weatherall2025some}. This, however, would not amount to a falsification of TEGR or STEGR. On the contrary, such a result could be seen as supporting them, as some proponents of TEGR have argued \citep[ch. 9]{aldrovandi2012teleparallel}. Specifically, such a finding would suggest that the SEP is not fundamental ($FEP$) but rather an emergent property ($EMEP$) arising from a possible gauge choice, consistent with TEGR and STEGR. In this case, it would mean that the SEP is recoverable only at limited levels, with constraints highlighted by experiments. This is analogous to Einstein's justification of the EEP on empirical grounds. In neither case do we have fundamental reasons to assume the given EP. However, GR explicitly assumes it as fundamental, with all the consequences we have discussed, whereas TEGR and STEGR do not. Moreover, recall that, as Wolf $\&$ Read argue, the empirical content of a theory is not fully determined by its dynamics \citep{Wolf_2023boundaries}, and theories sharing the same action may exhibit differences in their empirical content.

Another interesting possibility, suggested by the  Read and Teh taxonomy \citep{Read2023}, concerns the fact that a falsification of  WEP would not imply a falsification of SEP\(_{2,\mathrm{\neg EEP}} \), although its recovery should still be constrained by experimental data. In this scenario as well, TEGR would be a more fundamental theory than GR. Both these two scenario would be in line with the suggestion by Weatherall that both TEGR and STEGR have more surplus structure than GR \citep{weatherall2025some}.

However, it is important to emphasize that this argument encourages testing the  SEP but does not strictly predict its violation. In fact, even if the  SEP is not fundamental, it is still possible that no violation will be found. In such a case, increasing the accuracy of traditional EP tests may be insufficient to discriminate among the three theories.

This consideration suggests the need to explore and design other types of experiments that could highlight the  difference in \textit{essence} between gravity and inertia, investigate the possible distinct dynamics of the metric and the connection, or determine in other way whether the  SEP is fundamental or emergent. Consequently, these epistemic considerations point to the importance of developing experimental approaches beyond traditional EP tests, which might potentially differentiate among the three equivalent theories. This represents a subject for future research.

Then note that the continue corroboration of the  SEP at any level poses also other conceptual problems. If after significant progress, no violation will be found, at what level of accuracy would we consider it satisfied? $10^{-18}$? $10^{-20}$? $10^{-50}$? What level of precision could be deemed sufficient to corroborate a metric theory? This is a perfect example of inductive risk, as one could always find an experiment which violates the EP, even if we will reach a precision of $10^{-50}$. Is there an accuracy level where reasonable doubt would be mitigated? Maybe finding a more fundamental theory that explains why the  SEP should be an $FEP$ could help. However, justifying it experimentally seems difficult, as it is possible that its validity might turn out to be merely a contingent fact without any underlying fundamental reason. Philosophers of physics surely would enjoy the debate.\\

Finally, it is worth  mentioning another issue which could be addressed with our epistemic model, that is, the problem of the number of  degrees of freedom in theories of gravity \citep{magnano1994, Belenchia_2018, stabile2013}. The present approach could be exploited to extract  the number of degrees of freedom and the  dynamics of  competing theories, in order to evaluate  the equivalence among them and possible differences in their empirical content (see, e.g. \cite{Bajardi:2024dru,  Capozziello:2021pcg}).

\section{Conclusions}\label{Conclusions}
We discussed that  SEP is a non-trivial, theory-laden assumption in the framework of Equivalent Gravities. In fact, assuming  SEP at the foundation of the theory,  we are intrinsically stating  that gravitation is given by the curvature of spacetime, rather than by torsion or non-metricity. This assumption also leads to a number of other problematic foundational consequences: $i)$ the coincidence between the causal and the geodesic structure;  $ii)$ the fact that metric tensor is the fundamental variable of the theory instead of the connection; $iii)$ the contrast with several new physics predictions; $iv)$  the conceptual difficulties at quantum level; $v)$ the relation with CDM and DE problems; $vi)$ the conceptual difference of gravity with respect to  other gauge theories;  and, finally, $vii)$ the unjustified coincidence between  gravitational and  inertial mass. All these difficulties (or shortcomings) are direct or indirect consequences of the imposition of  SEP, which is based on the assumption that gravitation and inertia are of the same essence (as expressed by  EEP).

However,  GR is just a particular case in the more general lake of metric-affine theories, which can be built not only with curvature of spacetime, but also with torsion and non-metricity. With TEGR, built upon  torsion, and STEGR, built upon  non-metricity, GR constitutes the so called Geometric Trinity of Gravity, because these three theories are found to be dynamically equivalent.

Then, on a closer inspection, we argued that there is a crucial hidden difference  in  relation with  SEP. The significant fact is that, in both TEGR and STEGR,  SEP can be recovered but without the necessity to postulate it at the foundation of the theory. If physicists would have found the  SEP as a necessary fundamental principle also for TEGR and STEGR, this would have been considered as a meta-empirical evidence in favor of what we called the FEP, that is the Equivalence Principle as Fundamental. But, as we have seen, this is not the case. Therefore, given the equivalence among GR, TEGR and STEGR in non-trivial multiple predictions, and given the fact that   SEP is not necessary for TEGR and STEGR, our confidence in the fundamentality of the  SEP decreases.

If the argument is correct, this  structural divergence could lead to differences in the empirical content between the three theories, because if the  SEP is not a fundamental feature of reality (FEP), it is emergent (EMEP). And if it is emergent, it is possible that, at some level, it is not valid.

As argued, the relaxation of this theory-laden principle allows us also to address many of the aforementioned foundational problems. As Synge suggested, given its heuristic role, the EP could be considered as a midwife, but not a fundamental feature of the world \citep{Synge1960-SYNRTG}.

These epistemic considerations encourage physicists to further enhance the accuracy of EP tests, especially at the quantum level and unexplored regimes, and to develop new experimental schemes to investigate potential differences in the empirical content of theories arising from their distinct relationships with the EP, such as discriminating between dynamics of  metric and connection.


Finally, the approach of Sec. \ref{Epistemological considerations on Equivalent Gravities} could be used also for other investigations. First of all, it should be used to evaluate other predictions in order to search for other possible relevant observables. For example, principles  as the Local Lorentz Invariance, the Local Position Invariance, the Schiff conjecture, should be investigated also in TEGR and STEGR. Then, for instance, there could be also other equivalent formulations of gravity outside of the Geometric Trinity. Finding other dynamically equivalent theories would help in further  constraining the configuration space of the 
theory. Having already found two of such theories, nothing precludes the fact that  other equivalent representations of gravity could exist. Beside EP, theories of gravity could be compared also considering the number of degrees of freedom  related to observables.
This will be the argument of further studies.

\bmhead{Acknowledgements}
The Authors thank Elena Castellani, Carmen Ferrara, and Flaminia Giacomini for the useful discussions and feedbacks. 
C.M. and S.C. acknowledge the support of INFN sez. di Napoli, {\it iniziative specifiche} QGSKY and MOONLIGHT2. 
S.C. thanks the  {\it Gruppo Nazionale di Fisica Matematica} of {\it Istituto Nazionale di Alta Matematica} for the support.  This paper is based upon work from COST Action CA21136 - {\it Addressing observational tensions in cosmology with systematics and fundamental physics} (CosmoVerse), supported by COST (European Cooperation in Science and Technology).


\addcontentsline{toc}{chapter}{Bibliography}

\bibliography{sn-bibliography}{}


\end{document}